\documentclass{WileyMSP-template}
\usepackage{color}

\begin{document}

\pagestyle{fancy}
\lhead{\noindent{\textit{Advanced Intelligent Systems, Preprint Version, Published (July 2021)}}}
\rhead{\includegraphics[width=2.5cm]{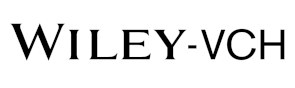}}

\title{Smart textiles that teach: Fabric-based haptic device improves the rate of motor learning}

\noindent \maketitle

\noindent 
\author{Vivek Ramachandran}
\author{Fabian Schilling}
\author{Amy R Wu}
\author{Dario Floreano*}



\begin{affiliations}
\noindent V. Ramachandran, F. Schilling, D. Floreano\\
School of Engineering, \'Ecole Polytechnique F\'ed\'erale de Lausanne,\\
1015 Lausanne, Switzerland\\
Email Address: dario.floreano@epfl.ch

\noindent A.R. Wu\\
Department of Mechanical and Materials Engineering, Queen's University\\
ON K7L 3N6 Kingston, Canada

\end{affiliations}


\noindent 

\begin{abstract}

\noindent People learn motor activities best when they are conscious of their errors and make a concerted effort to correct them.
While haptic interfaces can facilitate motor training, existing interfaces are often bulky and do not always ensure post-training skill retention.
Here, we describe a programmable haptic sleeve composed of textile-based electroadhesive clutches for skill acquisition and retention.
We show its functionality in a motor learning study where users control a drone's movement using elbow joint rotation.
Haptic feedback is used to restrain elbow motion and make users aware of their errors.
This helps users consciously learn to avoid errors from occurring.
While all subjects exhibited similar performance during the baseline phase of motor learning, those subjects who received haptic feedback from the haptic sleeve committed 23.5\% fewer errors than subjects in the control group during the evaluation phase.
The results show that the sleeve helps users retain and transfer motor skills better than visual feedback alone.
This work shows the potential for fabric-based haptic interfaces as a training aid for motor tasks in the fields of rehabilitation and teleoperation.

\end{abstract}



\section{Introduction}
Over the past decade, robotic teaching aids have been developed to train people in a variety of motor activities by providing sensory feedback \cite{johnson2006recent,sigrist2013augmented,reinkensmeyer2004robotics,lieberman2007tikl, marchal2009review, rauter2019robot, chan2010virtual}.
Motor learning is an error-driven process and the rate of learning depends on how sensory feedback is provided during training. 
Typically, people learn to rectify their errors based on a combination of visual, auditory, and haptic feedback \cite{sigrist2013augmented}.
Motor training by haptic feedback is of particular interest to researchers because it can be applied directly to the part of the body where corrective action is needed \cite{hirano2020overcoming, yin2020wearable}. 
This effectiveness hinges on two critical and intimately linked factors - the haptic interface and the training method \cite{marchal2010effect,mulder2012sharing,culbertson2018haptics}.

\noindent Haptic interfaces should provide reliable, intuitive, and clear feedback when required and be unobtrusive when they are not.
Within the scope of motor training, haptic interfaces are generally employed in two fields - rehabilitation and teleoperation \cite{marchal2009review,abbink2012haptic}. 
Existing haptic interfaces are mostly grounded i.e., they are fixed to a rigid base, which limits their applicability to tasks that do not require much user displacement, such as object manipulation \cite{gupta2006design}.
With the miniaturisation of electronic components, such as integrated circuits and batteries, wearable interfaces have become a viable alternative because they allow users a greater degree of mobility \cite{pacchierotti2017wearable}.
This increased mobility has been demonstrated through multiple exoskeletons for gait rehabilitation and robot teleoperation that improve user performance in the specific tasks for which they are designed~\cite{nef2007armin, klamroth2014three, duschau2010haptic, polygerinos2015soft,wehner2013lightweight, asbeck2015multi, low2017hybrid, asbeck2015biologically, panizzolo2016biologically, bimbo2017teleoperation,rognon2018haptic,rognon2019haptic}.
However, these interfaces are bulky because of the heavy actuators that they use, such as motors and pumps, which cause user fatigue over prolonged use. 
In fact, actuators are not always necessary to help users perform motor tasks better.
A recent study showed how a simple, unpowered clutch and spring ankle exoskeleton can increase human walking efficiency by re-purposing their expended energy \cite{collins2015reducing}.
Indeed, new lightweight, fabric-based haptic interfaces have been developed in recent years to circumvent the problems associated with using heavy actuators and promote user comfort for continuous usage \cite{liu2020textile, yin2020wearable, bianchi2014design, low2017hybrid, culbertson2018social, ramachandran2019all, park2019soft, wu2019wearable, carpenter2019healable, hinchet2020high, lee2020thermo, zhu2020pneusleeve} (Table \ref{table:table1}).
Amongst these fabric-based haptic interfaces, certain interfaces use electrostatic adhesive (EA) clutches to apply kinesthetic feedback through movement braking and passive springs \cite{hinchet2018dextres,hinchet2020high,ramachandran2019all,diller2016lightweight,diller2018effects}.
They operate at low power ($\sim$ \SI{1}{\milli \watt}) and are easily integrated into other textile-based wearables, such as clothing.
These new types of fabric-based, low-power-consuming haptic interfaces are less complex in mechanical design compared to existing interfaces, which are composed of rigid components.
The absence of actuators compels users to rely on the feedback to both identify and correct their errors, which helps them learn the motor task faster. 
Furthermore, unlike existing interfaces, these interfaces are not designed for any one specific task alone. Rather, they can be re-purposed to help train users in a variety of motor tasks.

        \begin{table*}
        \begin{center}
        \includegraphics[width=\textwidth]{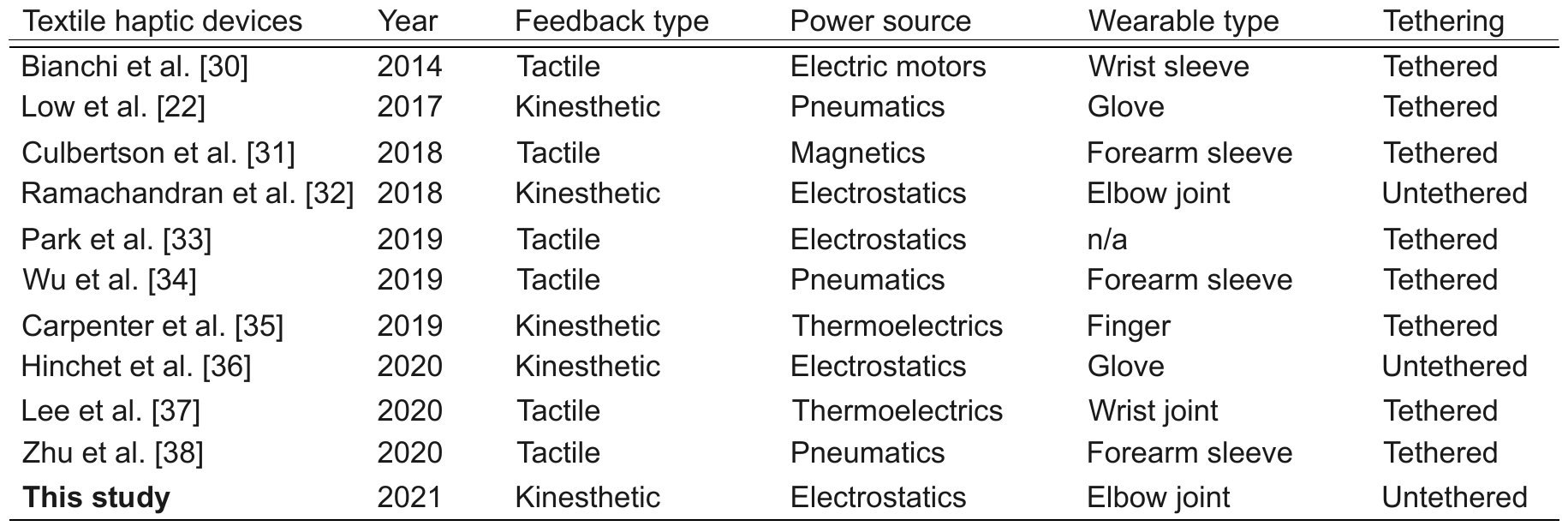}
        \caption{\textbf{Comparison of existing textile-based haptic devices and the one presented in this study}}
        \label{table:table1}
        \end{center}
        \end{table*}

\noindent A successful haptic training method ensures that users are immersed in learning the motor activity and are provided timely feedback to improve their performance over the course of training.
Existing haptic-based training methods can be broadly divided into two categories - haptic guidance and error amplification \cite{milot2010comparison,bouchard2015comparison}. 
In the former, the haptic system physically guides users to minimize errors they commit during training and accomplish a task.
In the latter, the haptic system amplifies user errors to intentionally increase the difficulty of the task \cite{marchal2013effect,lee2010effects}.
Studies show that haptic guidance increases user performance during training compared to baseline conditions, but that performance levels precipitate when the guidance is not provided \cite{powell2012task}.
Some have posited that the guidance overtly increases user dependency \cite{wulf2010frequent}.
This dependency curtails both skill retention and skill transfer post-training.
On the other hand, error-amplifying systems deliver longer periods of skill retention at the expense of longer training periods \cite{williams2016pays}.
However, some studies state that the comparative effects of haptic guidance and error amplification on motor learning cannot be generalised, because they are subject to the type of motor activity \cite{milot2010comparison,marchal2017effectiveness}.
Nonetheless, there is some consensus that the method of feedback provision and the resulting outcome is dependent on the user skill level \cite{marchal2010effect,marchal2017effect}.
Accordingly, novice users benefit more from haptic guidance, whereas expert users gain more from error amplification.
There is also recent evidence that suggests, allowing users to select the type and magnitude of haptic feedback can accelerate the rate of motor learning \cite{rauter2019robot, marchal2014optimizing}.
These studies conclusively show that users must actively utilize haptic feedback to rectify their errors and acquire motor skills.
However, so far, this active motor learning has only been demonstrated with haptic interfaces made with rigid and bulky devices \cite{rauter2019robot}.

\noindent Here, we present a novel haptic interface that promotes active user involvement in rectifying  movement errors during motor learning and is composed of only soft, fabric-based components.
The interface consists of a programmable elbow sleeve that comprises multiple fabric-based EA clutches, which can rapidly restrict the joint movements of a wearer.
In this study, the haptic sleeve is programmed to afford users a margin of error, but provides a motion-blocking feedback to 1) make users aware of their error, and 2) prevent these errors from growing.
The haptic feedback operates like a wall that confines the motion range.
Through heightened awareness of their errors, users can consciously avoid them in future iterations of the task.
In this study, we experimentally show that the proposed haptic interface increases the training success to learn, retain, and transfer motor skills in a drone teleoperation task. 

\section{Results}

\subsection{Design of the electroadhesive haptic sleeve}
\begin{figure*}
        \begin{center}
        \includegraphics[width=0.9\textwidth]{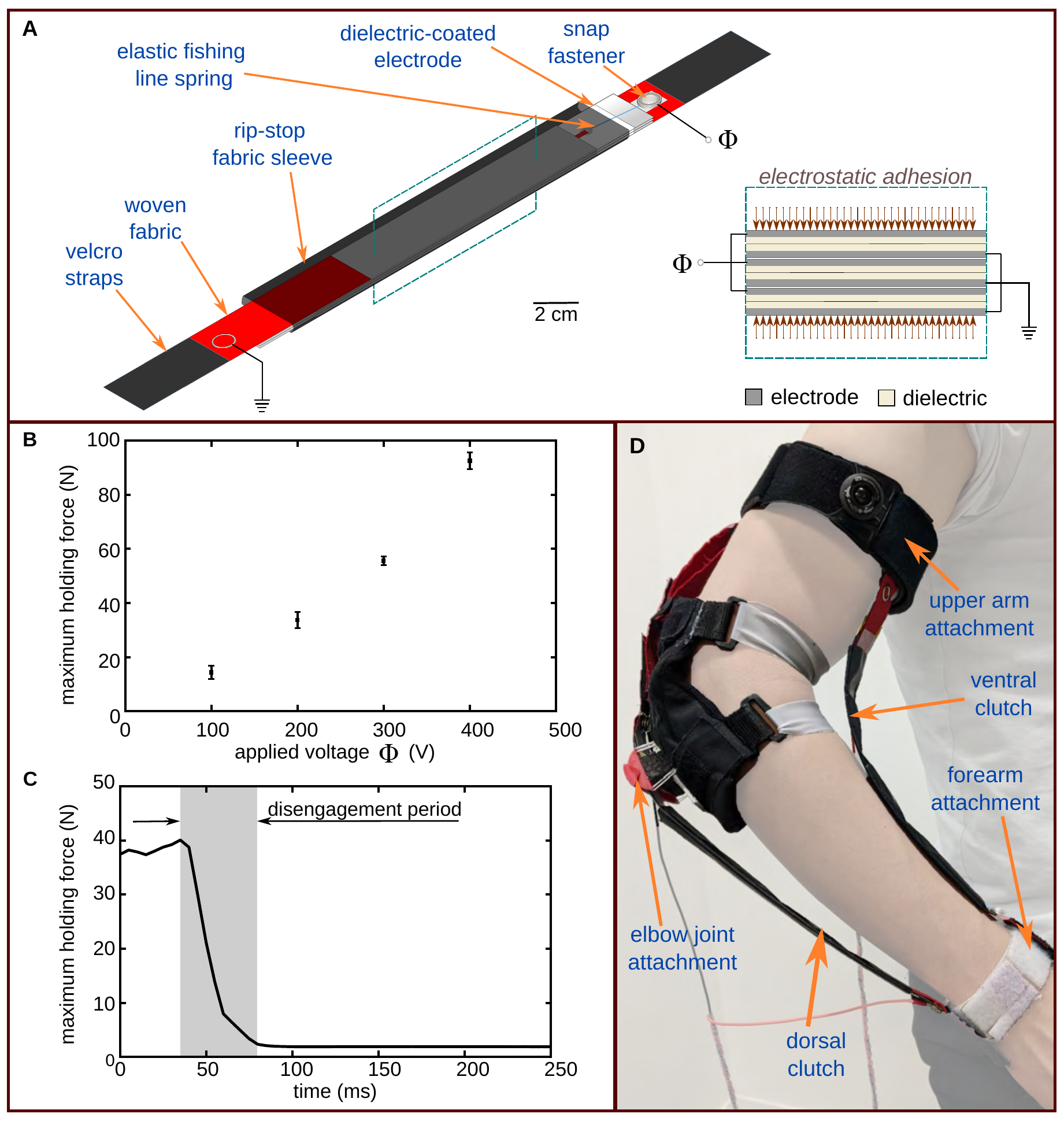}
        \caption{\textbf{Wearable haptic system to train users for motor activities.} (\textbf{A}) Design and composition of the fabric-based electroadhesive clutch. When a high voltage $\Phi$ is applied across the dielectric-coated electrode plates, the clutch engages and the clutch plates adhere to each other, thus increasing the maximum holding force. (\textbf{B}) The maximum holding force of the clutch is proportional to the applied voltage. Each data point is the mean value of five measurements trials and the error bars represent one standard deviation from this mean. (\textbf{C}) When the clutch is disengaged by removing the voltage, the adhesion decreases and the time required for the holding force to drop (gray shading) is measured as the disengagement time. (\textbf{D}) The wearable system is composed of two clutches, one each on the ventral and dorsal faces of the human arm. The clutches are held in place with the help of attachment straps on the forearm, the elbow joint, and the upper arm. The ventral clutch restricts elbow extension and the dorsal clutch restricts elbow flexion.}
        \label{fig:fig1}
        \end{center}
        \end{figure*}

The electroadhesive haptic sleeve is a fabric-based exoskeleton that can be programmed to constrain elbow extension and flexion.
It is composed of two electroadhesive clutches and three body attachments (Figure \ref{fig:fig1}A, Figure S1, Supporting Information).
Each clutch is a parallel plate capacitor composed of overlapping dielectric-coated electrodes. 
The EA clutches used in this study are an improved version of our previously described fabric-based clutches to generate higher holding forces\cite{ramachandran2019all}. 
The higher forces are generated by replacing the fabric-electrodes of our previous clutches with metallized biaxially-oriented polyethylene terephthalate sheets.
This was done to avoid dielectric cracking on the electrode surface, which was observed with the fabric electrode over prolonged usage.
Furthermore, surface irregularities and wrinkling were prevented by replacing the fabric electrodes with the metallized plastic sheets.
            
\noindent When a voltage in the order of \SI{100}{\volt} is applied across the overlapping electrode plates, they adhere to each other through electrostatic adhesion.
As a result, the maximum holding force i.e., the tensile force needed to ply the plates apart longitudinally increases. 
In this \textit{engaged} state, the magnitude of the holding force is dictated by the applied voltage (Figure \ref{fig:fig1}B).
When the clutch is \textit{disengaged} by removing the high voltage, it recovers its initial mechanical properties within \SI{40}{\milli\second} (Figure \ref{fig:fig1}C).
The elbow joint is a hinge-type joint.
Hence, the motion of the elbow joint can be blocked in a targeted manner by having one clutch each to block the flexion and extension of the forearm about the upper arm, without affecting the mobility of the remaining body.
To block forearm extension and flexion independently, one clutch is attached to the ventral face of the arm and another to the dorsal face, respectively.
Each clutch can be extended longitudinally from its rest length by virtue of low stiffness springs, which ensure that the user does not experience any hindrance to natural mobility when haptic feedback is not applied.
When the voltage is applied, the electrostatic adhesion prevents further longitudinal extension of the clutch and thus constrains joint rotation.
The perception of the haptic feedback is quantified by the magnitude of the holding force of the EA clutches.
This magnitude is dependent on the dimension of the clutch plates, the dielectric thickness, the number of the clutch pairs and the applied voltage.
Perception of the haptic devices can be tuned by changing any of these parameters.
The ventral and dorsal clutches are worn by placing them between body attachments that are anchored to the forearm, the elbow joint, and the upper arm (Figure \ref{fig:fig1}D, Movie S1, Supporting Information).
The forearm and the upper arm attachments are made of long strips of elastane bonded to loop fastener strips (Velcro) on the exterior and a layer of silicone (Ecoflex 00-20, Smooth-On) on the interior (Figure S2, Supporting Information).
Once the clutches are mounted, a custom dial strap with a BOA tightening system (3M ACE Brand, MN, USA) is worn over the upper arm attachment to secure the anchoring. 
The manufacturing process of the forearm and upper arm attachments involves bonding the elastane to a membrane of cured Ecoflex 00-20 by first, casting an uncured layer of Ecoflex 00-20 on the elastane, then, placing the cured membrane on top of the uncured layer and finally, oven-curing the entire composite.
This part of the process was adapted from earlier work carried out by colleagues \cite{tonazzini2018variable}.
The Ecoflex is comfortable and adheres to human skin.
The loop fastener on the body attachment mates with the hook fastener of each clutch.
The elbow joint attachment is necessary for the functioning of the dorsal clutch.
A semi-cylindrical 3D printed guide is sewn onto the pad at the elbow joint to allow for longitudinal extension of the dorsal clutch (Figure S3, Supporting Information).
Two inertial measurement units (IMU, by Xsens Technologies) are attached to the forearm and upper arm body attachments to measure the elbow angle.

\subsection{Motor learning to teleoperate a drone} 
\begin{figure*}
        \begin{center}
        \includegraphics[scale = 0.82]{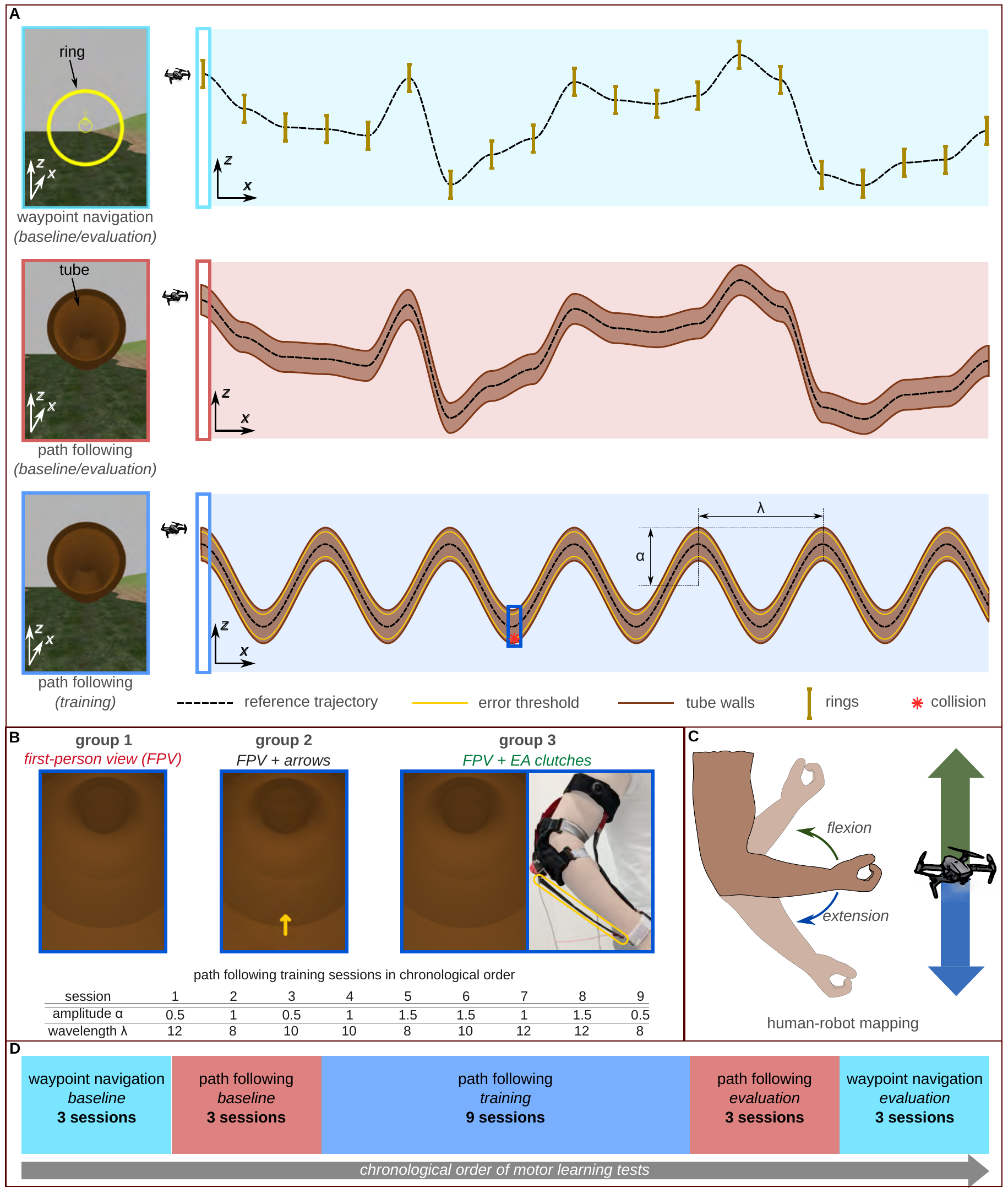}
        \caption{\textbf{Experimental pipeline for motor learning} (\textbf{A}) Subjects performed two motor tasks - waypoint navigation (rings) and path following (tubes). Cubic splines produced reference trajectories for baseline and evaluation phases of both tasks while sine curves defined by amplitude ($\alpha$) and wavelength ($\lambda$) were used for the training phase of path following. (\textbf{B}) Subjects in each group received differentiated feedback when they traversed an error threshold within the training phase tubes:  first-person view (FPV),  FPV + arrows, or FPV + EA clutches. The training sessions are listed chronologically with the $\alpha$ and $\lambda$ values. (\textbf{C}) A linear mapping was made between the subject's elbow angle and the drone's altitude. (\textbf{D}) The experimental pipeline is shown in chronological order.}
        \label{fig:fig2}
        \end{center}
        \end{figure*}

We assessed the applicability of the EA haptic sleeve as a teaching aid for drone teleoperation tasks and used user errors as the performance metric to determine skill retention and transfer.
The experiments consisted of two drone teleoperation tasks - path following to examine the effect of haptic training on the retention of motor skills, and waypoint navigation to determine the transfer of those skills (Figure \ref{fig:fig2}A).
Both tasks are frequently used methods in the context of drone teleoperation. 
Path following is often employed to navigate drones through long, narrow pipelines for maintenance and inspection or navigate through cluttered environments that are inaccessible to humans.   
Waypoint navigation is commonly used to help drones map areas of interest or carry out aerial monitoring \cite{otto2018optimization}.
For the path following motor task, subjects were asked to control the altitude of a drone flying through a cylindrical tube with multiple vertical bends while avoiding collisions with the walls.
For the waypoint navigation motor task, subjects were asked to control the altitude of a drone through a series of rings  positioned at different heights. 
The performance of the path following task was computed by measuring the altitude error with respect to the desired centreline of the tube throughout its length.
The performance of the waypoint navigation task was computed by measuring the difference in height between the drone and the centre of each traversed ring.
Waypoint navigation is chosen for the skill transfer test because it is similar to path following.
Greater task similarity between the skill retention and transfer tests improves the chances of skill transfer.
Despite their similarities though, waypoint navigation provides users more freedom in terms of drone movement than path following as the performance is only evaluated at discrete intervals.
However, this additional freedom comes at the risk of increased movement errors if the user does not retain and transfer the necessary skills to control the drone.

\noindent There is a linear mapping between the elbow angle measured by the IMUs and the altitude of the simulated drone (Figure \ref{fig:fig2}C, Movie S2, Supporting Information).
The IMU readings of the elbow angle are continuous i.e., they are non-discrete.
The amplitude limits are determined by initially measuring the elbow angles for each subject at maximum forearm flexion and extension and then, scaling the altitude limits of the drone in the simulated environment to match the elbow angle limits.
The path following task consisted of three phases - baseline, training, and evaluation, while the waypoint navigation tasks consisted of two phases - baseline and evaluation.
In chronological order, subjects performed the baseline phase of waypoint navigation, the baseline, training, and evaluation phases of path following, and finally, the evaluation phase of waypoint navigation (Figure \ref{fig:fig2}D). 
The baseline and evaluation phases of both the waypoint navigation and path following tasks consisted of three sessions each, during which subjects received only visual feedback from the drone's camera in first-person view (FPV) via a desktop display (Movie S3, Movie S4, Supporting Information).
Cubic splines were used to produce the reference trajectories for the baseline and evaluation phases of both tasks.
The tubes for the training phase of the path following task were produced with sine curve reference trajectories.
This is due to the periodic nature of sine curves, which allowed participants to become acclimatized to the nature of the experiment.
Since each sine curve is defined by a specific combination of an amplitude and a wavelength, three amplitudes $\alpha$ (\SI{0.5}{\meter}, \SI{1}{\meter}, and \SI{1.5}{\meter}) and three wavelengths $\lambda$ (\SI{8}{\meter}, \SI{10}{\meter}, and \SI{12}{\meter}) were used to produce nine distinct tubes.
Prior to the commencement of the human subject studies, the chronology of the training sessions was set by randomizing the order of the amplitude and wavelength combinations (Figure \ref{fig:fig2}B).
This was done to prevent biasing effects due to amplitude or wavelength.
This randomization was only performed once, and the chronological order was maintained for all subjects.
During the training phase, subjects received additional sensory feedback and were grouped accordingly: group 1 - FPV; group 2 - FPV with yellow arrows indicating corrective action; and group 3 - FPV  with EA clutches providing haptic feedback to block elbow joint rotation (Figure \ref{fig:fig2}B, Movie S5, Movie S6, Movie S7, Supporting Information).
An error threshold of \SI{0.4}{\meter} was set above and below the centreline of the training tubes, which was not visible to the subject.
Whenever subjects in groups 2 and 3 flew the drone past this threshold, they received the additional sensory feedback (arrows/EA clutches) to prevent them from colliding with the tube wall.
For group 2, the arrows pointed downward when the drone was close to the ceiling and they pointed upward when the drone was close to the floor.
Similarly, for group 3, the dorsal clutch blocked elbow flexion to prevent ceiling collisions and the ventral clutch blocked elbow extension to prevent floor collisions.

\subsection{Acquisition and retention of motor skills} 
\begin{figure*}
        \begin{center}
        \includegraphics[width = \textwidth]{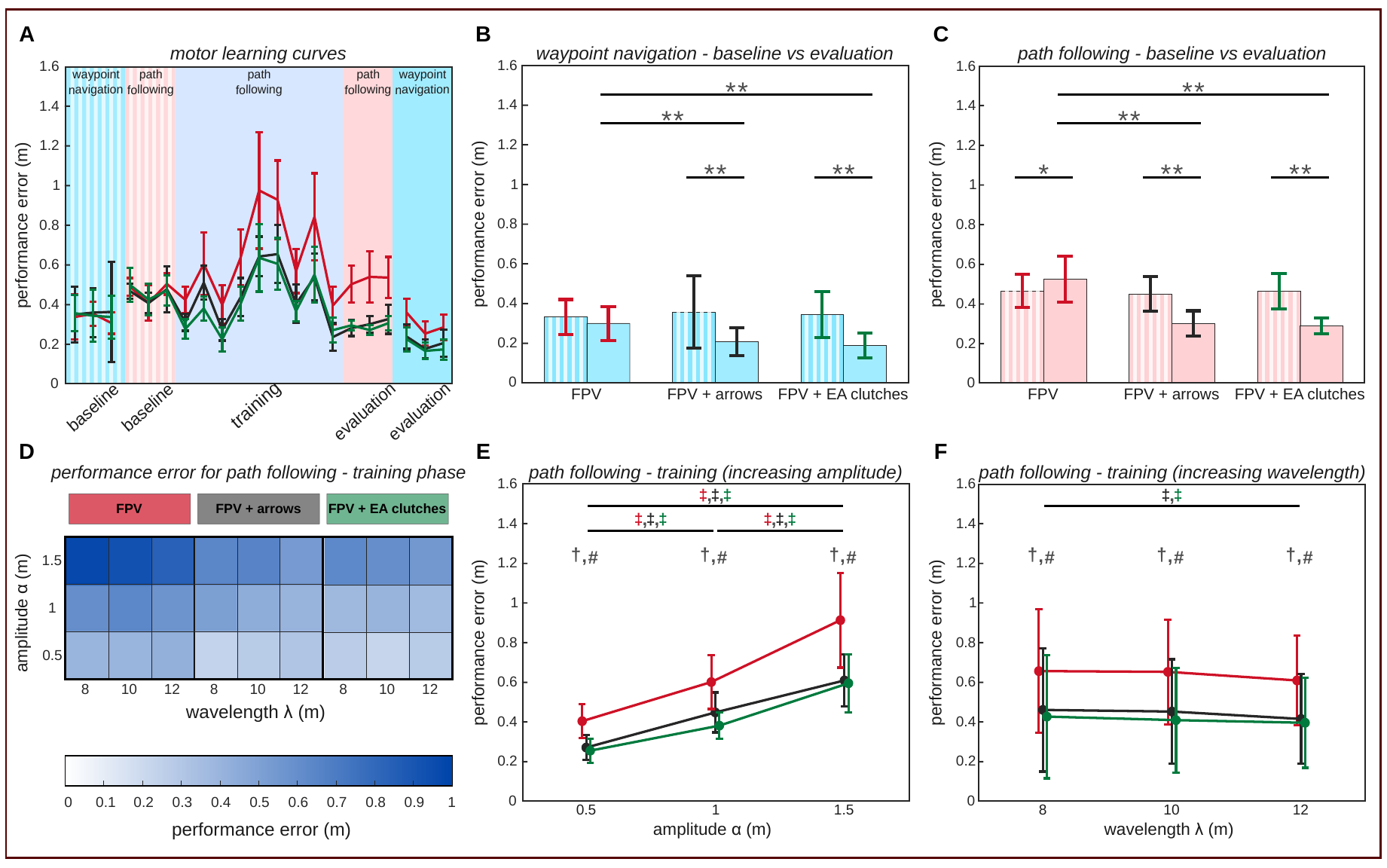}
        \caption['long']{\textbf{Results from the motor learning human subject study.} (\textbf{A}) Motor learning curves of each subject group are plotted as a function of mean performance error against experimental phases of both motor tasks in chronological order. Subject groups: group 1 (FPV, red), group 2 (FPV + arrows, black), group 3 (FPV + EA clutches, green); waypoint navigation task - baseline (light blue, striped) and evaluation (light blue, solid) phases; path following - baseline (pink, striped), training (light purple, solid), and evaluation (pink, solid) phases. (\textbf{B})-(\textbf{C}) Between-group and within-group comparisons of performance errors during baseline and evaluation phases of waypoint navigation and path following tasks, *P$<$0.05, **P$<$0.01. (\textbf{D}) Heat map of group performance errors during the training phase of path following shows the effects of amplitude ($\alpha$) and wavelength ($\lambda$) combinations. (\textbf{E})-(\textbf{F}) Comparisons between-group and within-group of subject performance during the training phase of path following. Each plotted point is the average performance error of a group for three training sessions with the same amplitude (\textbf{E}) or wavelength (\textbf{F}). $\dagger$ Comparison between group 1 and group 3, P$<$0.01.  \# Comparison between group 1 and group 2, P$<$0.01. $\ddagger$ Within-group comparison between different amplitudes or wavelengths colour-coded to match groups, P$<$0.01.  All error bars in this figure show one standard deviation above and below the reported mean performance error.}
        \label{fig:fig3}
        \end{center}
        \end{figure*}

Subject performance errors were measured for both teleoperation tasks and the temporal changes in their performance described the motor learning characteristics of each group (Figure \ref{fig:fig3}A).
Differences in performance errors between groups for the same task phases revealed the effects of differentiated feedback provision.
The performance errors of the path following task were compared to understand how each of the groups attempted to acquire and retain motor skills specific to this task (Figure \ref{fig:fig3}C).
Subjects in groups 2 and 3 committed fewer errors (32.89\% and 37.56\% respectively) in the evaluation phase compared to the baseline phase for the path following task.
On the other hand, subjects in group 1 committed  12.65\% more errors during the evaluation phase of path following compared to the baseline phase.
A paired \textit{t}-test showed that the performance of subjects in group 1 (FPV) deteriorated from baseline to evaluation for the path following task indicating that the training phase between the baseline and evaluation phases did not benefit the subjects and may have even been deleterious to them ($t=$-2.535, $P<$0.05).
On the other hand, subjects in group 2 (FPV + arrows) and group 3 (FPV + EA clutches) performed significantly better in the evaluation phase compared to the baseline, which implies that they were able to learn the task over the course of training (group 2: $t$=23.76, $P<$0.01; group 3: $t$=10.041, $P<$0.01).
Indeed, a one-way analysis of variance (ANOVA) followed by a \textit{post-hoc} Holm-Sidak corrected \textit{t}-test showed that there were no significant differences between the performances of the three groups at baseline ($F_{2,27}$=0.34, $P$=0.71).
This outcome ruled out any biases that might have been introduced by relative differences in subject task expertise prior to training.
On the other hand, the same statistical treatment revealed that the performance errors of subjects in group 2 and group 3 were lower than those of subjects in group 1 after evaluation, thus showing the retention of motor skills after training ($F_{2,27}$=84.93, $P<$0.01).
During the training phase of path following, the subjects of each group received differentiated feedback as explained earlier.
In addition to the received sensory feedback, the performance of each group were subjected to the effects of different amplitude and wavelength combinations that defined each tube's reference trajectory (Figure \ref{fig:fig3}D).  
For each group, these effects were resolved by averaging their performance errors of a set of three training sessions.
Each set had the same amplitude (\SI{0.5}{\meter} - sessions 1, 3, 9; \SI{1}{\meter} - sessions 2, 4, 7; \SI{1.5}{\meter} - sessions 5, 6, 8; Figure \ref{fig:fig3}E) or the same wavelength (\SI{8}{\meter} - sessions 2, 5, 9; \SI{10}{\meter} - sessions 3, 4, 6; \SI{12}{\meter} - sessions 1, 7, 8; Figure \ref{fig:fig3}F).
For a given amplitude or wavelength, each comparison was made between groups using a one-way ANOVA followed by a \textit{post-hoc} Holm-Sidak corrected \textit{t}-test.
Groups 2 and 3, which received augmented visual and haptic feedback, respectively, performed significantly better than group 1 for each training set grouped by amplitude ($\alpha$=0.5, $F_{2,27}$=20.64, $P<$0.01; $\alpha$=1, $F_{2,27}$=16.6, $P<$0.01; $\alpha$=1.5, $F_{2,27}$=12.33, $P<$0.01).
Similar results were obtained  when comparing the performance errors of groups 2 and 3 with respect to group 1 for training sets grouped by wavelength ($\lambda$ = 8, $F_{2,27}$=14.75, $P<$0.01; $\lambda$ = 10, $F_{2,27}$=23.5, $P<$0.01; $\lambda$ = 12, $F_{2,27}$=22.18, $P<$0.01).
These two results pertaining to the amplitude and wavelength effects on group performance errors showed that the additional sensory feedback alerted subjects in groups 2 and 3 in time to switch between extension and flexion before colliding with the tube walls.
A repeated measured ANOVA followed by a \textit{post-hoc} Holm-Sidak corrected \textit{t}-test showed that subjects in each group performed progressively worse as the amplitude of the tube reference trajectory increased with observable significant differences (group 1: $F_{2,18}$=40.71, $P<$0.01; group 2: $F_{2,18}$=144.14, $P<$0.01; group 3: $F_{2,18}$=50.75, $P<$0.01).
This is expected given the increasing difficulty to alternate between forearm extension and flexion at larger elbow angles.
The \textit{post-hoc} Holm-Sidak corrected \textit{t}-test values for pairwise comparisons within-group for different amplitudes are reported (Data file S1, Supporting Information). 
The repeated measures ANOVA did not reveal significant differences in performance for group 1 for varying wavelengths ($F_{2,18}$=2.74; $P$=0.09).
However, a repeated measures ANOVA followed by  \textit{post-hoc} Holm-Sidak corrected \textit{t}-test revealed that groups 2 and 3 performed significantly better for a wavelength of \SI{12}{\meter} compared to \SI{8}{\meter} (group 2: $F_{2,18}$=4.93, $P$=0.01; group 3: $F_{2,18}$=4.71, $P$=0.02).
The less marked differences in wavelength effects within each group was perhaps due to the chosen drone forward speed.
Different drone speeds might have produced more noticeable differences.

\subsection{Transfer of motor skills}
Subject had to utilize the motor skills garnered during the path following task to improve upon their baseline performance in the waypoint navigation transfer task (Figure \ref{fig:fig3}B).
A one-way ANOVA with \textit{post-hoc} Holm-Sidak \textit{t}-test corrections established no significant differences between the baseline performances of each group, thus reinforcing the absence of expertise bias as shown for the baseline phase results of path following ($F_{2,27}$=0.12, $P$=0.88).
While all subjects in each group performed better during the evaluation phase of the waypoint navigation task compared to their baseline, the relative improvement in performance was more pronounced for subjects in group 2 and 3, where the error percentage drop was 42.23\% and 45.64\% respectively, compared to the 9.97\% drop for subjects in group 1.
A paired \textit{t}-test showed that the performance errors did not change significantly for subjects in group 1 between the baseline and evaluation phases ($t$=1.27, $P$=0.23).
This may be attributed to inadequacies in motor skills acquired during the path following task.
In the same vein, subjects in groups 2 and 3, who exhibited performance improvement for the path following task, produced similar results for waypoint navigation.
There were significant differences between their baseline and evaluation phase performances showing that the skills acquired from path following transferred (group 2: $t$=3.82, $P<$0.01; group 3: $t$=7.31, $P<$0.01).
A one-way ANOVA with \textit{post-hoc} Holm-Sidak \textit{t}-test corrections showed that subjects in groups 2 and 3 performed significantly better during the evaluation phase compared to group 1, thereby showing the relative benefits of having been trained with additional sensory feedback ($F_{2,18}$=0.12, $P<$0.01).

        \begin{figure*}
        \begin{center}
        \includegraphics[width = \textwidth]{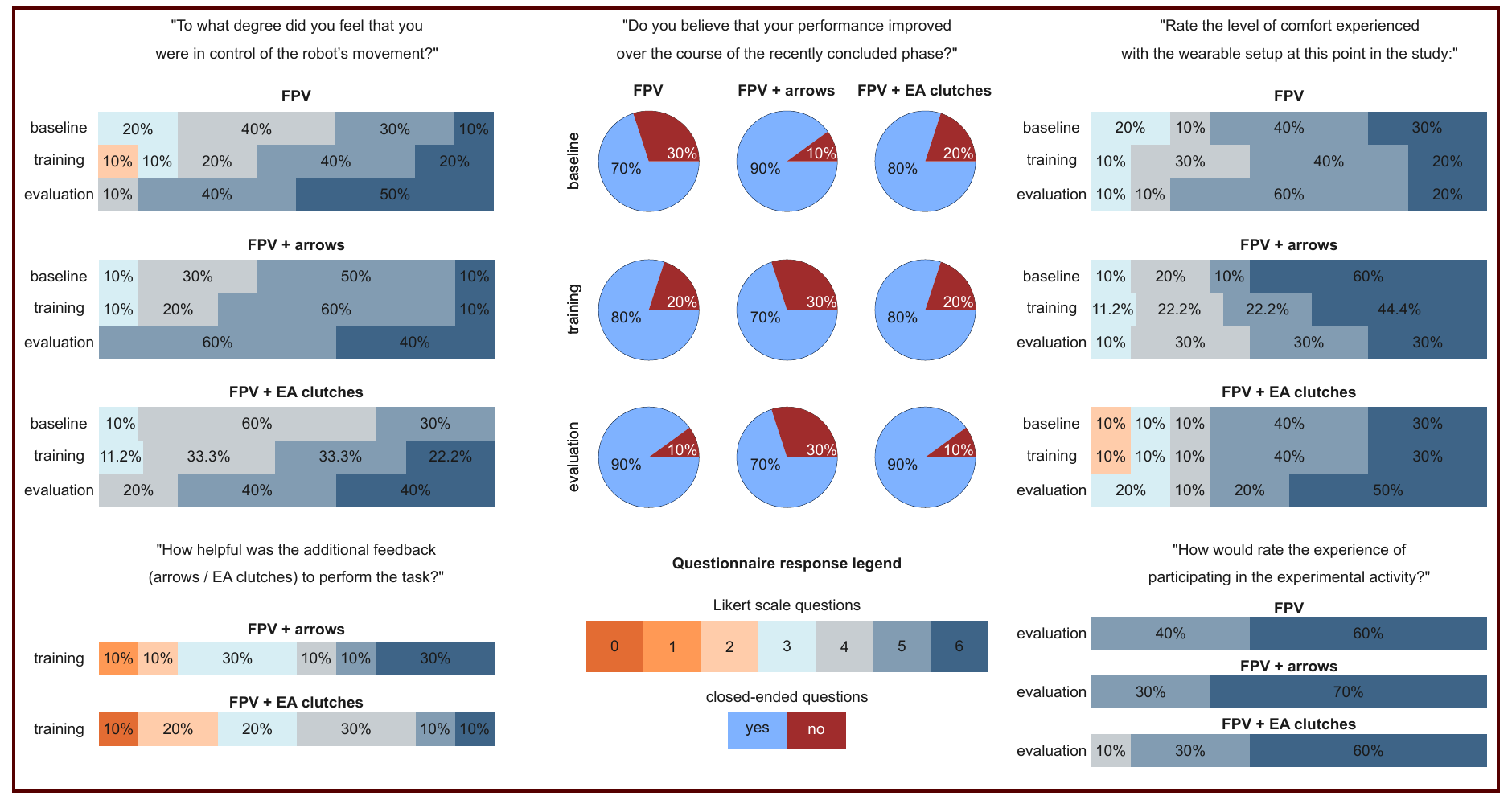}
        \caption['long']{Subject responses to questionnaires filled after the baseline, training, and evaluation phases for path following. Questions posed determined various qualitative aspects of their motor training, including their comfort levels with both the wearable interface and the control of the drone control, as well as their own assessment of performance improvement at the end of each of these phases. At the end of the training phase, subjects in the test groups that received additional feedback in the form of corrective arrows or EA clutches rated the degree to which this feedback helped them over the course of the phase. Finally, subjects provided an overall rating of their experience in participating in the motor training exercise. With the exception of the yes/no response to the question about their performance self-assessment, responses to all other questions were provided on a 7-point Likert scale between 0 and 6.}
        \label{fig:fig4}
        \end{center}
        \end{figure*}

\subsection{Subjective assessment of drone teleoperation and sleeve comfort}
We assessed each subject's comfort in learning to control the drone while wearing the haptic sleeve using questionnaires. 
Since differentiated feedback was only provided during the path following phase, subjects filled three questionnaires, each corresponding to the three phases of path following -  \textit{baseline}, \textit{training},  and \textit{evaluation} at the conclusion of each respective phase.
Subjects were asked to assess the degree to which they were in control of the drone's movement, their level of comfort with the haptic interface, and a self-evaluation of their performance improvement over the recently concluded phase by indicating a grade on a 7-point Likert scale (0-6).
For all groups, at least \SI{80}{\percent} of the subjects rated both their degree of control over the drone's movement and their level of comfort with the wearable system between 4 and 6 (both included) for all three phases (Figure \ref{fig:fig4}).
Subjects of groups 2 and 3 were asked to rate the helpfulness of the additional sensory feedback (arrows and EA clutches) in performing the tasks as part of the training questionnaire.
Despite the quantitative results which indicated that subjects in these two groups improved upon their baseline performance, only \SI{50}{\percent} of them found the additional sensory feedback (arrows or EA clutches) to be qualitatively helpful.
It is interesting that participants do not perceive the feedback as qualitatively helpful and yet, they tend to perform better. 
The rather counter intuitive response from the subjects may be due to the additional feedback creating sensory overload, but this claim cannot be verified as part of our study.
For the self-evaluation of performance improvement subjects could answer ``Yes" or ``No".
For group 1, the percentage of subjects who responded to their self-evaluation of performance improvement as ``Yes" increased (70\% - ``Baseline"; 80\% - ``Training"; 90\% - ``Evaluation").
On the other hand, the percentage reduced for group 2 (90\% - ``Baseline"; 70\% - ``Training" and ``Evaluation").
The percentage of affirmative responses increased for group 3 over the phases (80\% - ``Baseline" and ``Training"; 90\% - ``Evaluation").
In the evaluation questionnaire for all groups, subjects were also asked to rate their overall experience in participating in the experimental study. 
All subjects rated their overall experience between 4 and 6 in participating in the experiments.

\section{Conclusion}
We described a wearable haptic sleeve that uses fabric-based EA clutches to impart kinesthetic feedback by blocking body joint movement and experimentally showed its functionality as a teaching aid for motor activities in drone teleoperation tasks.
This study examined and compared the effects of providing haptic feedback for motor training with different forms of visual feedback.  

\noindent The results show that subjects in the control group, who received FPV visual feedback did not acquire and retain sufficient motor skills to improve path following performance and subsequently, were unable to perform better in the transfer waypoint navigation task.
Instead, subjects who received either augmented visual feedback in the form of arrows to correct elbow movement or haptic feedback from the electroadhesive haptic sleeve to physically block elbow rotation, displayed performance improvement from baseline to evaluation for both path following and waypoint navigation.
This performance improvement could be attributed to users relying on both types of feedback to determine their errors and consciously learning to to avoid them.
This conclusion is reinforced by the observable improvement in performance level with respect to the control group immediately after receiving additional haptic/visual feedback.
While this additional feedback assists subjects in identifying and avoiding errors during the training phase, their newly acquired motor skills  are retained even after the training phase.
This show that the subjects did not become overtly dependent on the additional feedback.

\noindent The increase in performance errors with the amplitude of tube reference trajectories for all subject groups can be ascribed to the difficulty in rapidly alternating between forearm extension and flexion, especially when approaching the maximum and minimum elbow angle limits.
Indeed, the performance errors of specific training sessions on average were higher for all groups than individual sessions within the baseline and evaluation phase due to the training tubes having larger amplitudes.
The relatively small differences in performance for all groups with changing tube wavelength may be due to the chosen drone speed.
For higher drone speeds, we might expect to see greater performance differences with fewer errors being committed for larger tube wavelengths.

\noindent The comparable beneficial effects of haptic feedback and of corrective arrow displays are noteworthy because augmented visual feedback is generally accepted as a benchmark in feedback-based training. 
Indeed, there are no observable statistically significant differences between the two forms of additional feedback based on the data collected during the subject studies.
While the haptic feedback is just as effective as a teaching aid as augmented visual feedback, haptic feedback could be qualitatively more helpful than augmented visual feedback when provided directly to the part of the body responsible for erroneous motion.
This is because concentrating the relay of augmented sensory feedback through a single feedback channel (vision) can severely increase the risk of sensory overload over prolonged periods of robot teleoperation \cite{walker2020continuous}.
Furthermore, haptic feedback could be used instead of vision when visual feedback from the robot is occluded \cite{walker2020continuous, macchini2020hand}.
Certainly, visual occlusions are not uncommon when operators need to inspect infrastructure, for example maintenance after the occurrence of a natural disaster.
In these instances, first-person visual perspective may not reveal concealed structures in the robot's periphery due to a variety of reasons, including insufficient lighting. 
Another possibility is that visual feedback may be cut off from the operator intermittently due to poor transmission.
In addition, haptic feedback can be provided to the visually impaired.
In this work, the haptic feedback was binary i.e., either it behaved as a compliant cloth or it blocked the range of motion.
In the future, lightweight variable stiffness technologies which use low melting point materials and shape memory materials that can provide a range of blocking forces could also be employed for different training tasks.
Furthermore, extension of this technology to other joints, such as the wrist and fingers, could be used for more complex teleoperation and rehabilitation tasks.

\section{Experimental Section}

\textit{Manufacturing of electroadhesive haptic clutches}

\noindent Each clutch consisted of three pairs of dielectric-coated electrodes that were interleaved in an interdigitated architecture.
The electrodes (\SI{150}{\milli\meter} $\times$ \SI{35}{\milli\meter}) were \SI{15}{\micro\meter}  biaxially-oriented polyethylene terephthalate that were metallised on one surface.
The metallised surface was coated  with a \SI{20}{\micro\meter} layer of high- $\kappa$ dielectric ink, a ferroelectric composition of Barium Titanate and Titanium Dioxide (Luxprint 8153, DuPont). 
The dielectric was oven-cured at \SI{140}{\celsius} for \SI{60}{\min}.
Post-curing, the solid dielectric that remained was \SI{10}{\micro\meter} thick.
The non-metallised surface of each electrode was bonded to a \SI{120}{\micro\meter} sheet of polymethyl methacrylate (PMMA).  
Each of the three pairs of interleaved electrodes overlapped with their dielectric surfaces in contact.
By virtue of the interdigitated architecture, one electrode of each pair was maintained at high voltage with respect to its paired electrode that was grounded when the clutch was engaged.
Therefore, the two sets (high voltage and ground) of three electrodes that were maintained at the same voltage were bonded together using electrically conductive copper‐plated polyester fabric tape (CN‐3190, 3M). 
Each set was adhered to a strip of woven fabric.
\SI{50}{\milli\meter} long silicon springs (Zim Fluo \SI{0.6}{\milli\meter} diameter elastic fishing line, Autain Peche) were attached between one electrode set and the woven fabric strip adhered to the other electrode set via bolted snap fasteners.
A sleeve of rip-stop fabric was used to ensure that the electrodes plates do not move laterally.
Hook fastener (Velcro) strips are stitched onto the woven fabric to mount the clutches to the forearm and upper arm body attachments. 
The clutches are operated by a customized printed circuit board that can apply high voltages when serial commands are sent via Bluetooth from the computer running the motor learning simulations.
The net weight of each clutch was \SI{25}{\gram}.
\linebreak

\noindent \textit{Mechanical and electrical characterisation of clutches}

\noindent The maximum holding force of each clutch was measured by performing force-displacement tests with a materials testing machine Instron 5965 (Instron, Norwood, MA, USA).
The ends of the clutch were fixed to the vices of the tensile tester. 
Each of the three clutch pairs have an initial dielectric overlap area of 120 $\times$ \SI{35}{\milli\meter\squared}. 
The clutch was engaged at different voltages (\SI{100}{\volt}, \SI{200}{\volt}, \SI{300}{\volt}, and \SI{400}{\volt}) and loaded in tension at a rate of \SI{10}{\milli\meter\per\second}.
The maximum holding force was reported for each voltage as an average of five measurement trials. 

\noindent To determine the clutch disengagement time, the clutch was initially fixed between the vices of the tensile tester and then engaged at \SI{250}{\volt}.
The clutch was loaded in tension until it reached the maximum holding force at which point the clutch is disengaged by setting the applied voltage from the board to zero.
At the same time, a serial command was sent to the tensile tester to register the time of disengagement. 
The disengagement characteristics were reported by averaging the precipitating holding force measurements for three trials.
The disengagement time was measured by computing the \SI{90}{\percent} drop from the initially measured maximum holding force corresponding to the applied voltage. 
\linebreak

\noindent \textit{Simulation framework for drone teleoperation}
\noindent The drone flight tasks were simulated using Gazebo, an open source robotic simulator \cite{koenig2004design}.
Gazebo provided 3D visual scene rendering and simulated on-board sensors such as RGB cameras.
The simulation environment and drone model were based on our previous work on machine-learning-based multi-drone control \cite{schilling2019learning}. 
For the flight tasks, single-integrator dynamics were used to move the drone through the environment.
The drone flight was restricted within the x-z plane and moved forward at a constant speed of \SI{5}{\meter\per\second}. 
For each of the three baseline and evaluation sessions of waypoint navigation, a set of 20 rings of 1m diameter each were generated in the environment.
The rings were spaced at equal intervals of \SI{4}{\meter} along the x-axis with the openings facing the drone's direction of travel.
For each session, the reference trajectory passing through the ring centres was created using cubic spline, such that spline slope was zero at the ring centres.
Furthermore, the z-position of the ring centres were constrained to remain between \SI{9}{\meter} and \SI{11}{\meter} above the ground. 
For the baseline and evaluation sessions of path following, \SI{1}{\meter} diameter tubes measuring \SI{76}{\meter} in horizontal length were rendered in the same x-z plane as the drone, appearing \SI{4}{\meter} in front of it.
The centreline of each tube, which was the drone's reference trajectory was also produced using cubic splines.
The trough and the peak of the trajectory were constrained to \SI{9}{\meter} and \SI{11}{\meter} respectively. 
For the training sessions, 1m diameter tubes measuring \SI{76}{\meter} in horizontal length were produced with tube centreline reference trajectories as sine curves.
To generate nine distinct tubes, nine distinct combinations of amplitude and wavelength were used.
The three amplitudes values were \SI{0.5}{\meter}, \SI{1}{\meter}, and \SI{1.5}{\meter}, and the three wavelength values were \SI{8}{\meter}, \SI{10}{\meter}, and \SI{12}{\meter}.
\linebreak

\noindent \textit{Human subject study}

\noindent A total of 30 adult subjects (ages between 23 and 42, mean = 29.86, standard deviation = 4.32) were recruited, primarily from the university.
Each of the subjects were healthy, had normal hearing and normal or corrected-to-normal vision.
None of the subjects had prior knowledge of the tasks they were expected to perform.
They were randomly assigned to the three groups with 10 subjects each (6 men, 4 women in each group).
The subjects provided written informed consent, and the study was approved by the EPFL Human Research Ethics Commission.
Before the motor learning studies were performed, preliminary psychophysical measurements were made to ascertain the Just Noticeable Difference in terms of the holding force needed to constrain their elbow flexion and extension.
The applied DC voltage was directly proportional to the maximum holding force.
A voltage value of \SI{300}{\volt} corresponding to the results of the Just Noticeable Difference was used as the operating voltage for the engagement of the clutches during the experiments.
The dimensions of the haptic devices were determined  \textit{a priori} by using the anthropometric data collected from 10 individuals, specifically, their forearm length, their upper arm length, and the changes in length on the ventral and dorsal arm faces associated with forearm extension and flexion about the upper arm respectively.
While the haptic devices are robust to repeated usage, the elastic fishing line used as the low-stiffness spring can become brittle at the knots, as is often the case with some types of vulcanized rubber. 
To prevent any deleterious effect as a result of this changing mechanical property, the springs were replaced after every 5 subjects.
The clutch plates were operated at a voltage that would not cause electrical shorting.
To prevent the accumulation of surface charges on each clutch plate due to the DC voltage operation, the polarity of the clutch plates was reversed for each subject after every session of each experimental phase.
\linebreak

\noindent \textit{Statistics}

\noindent All analysis on the data gathered from the simulated flight tasks was carried out in MATLAB 2019b (Mathworks, MA, US). 
The paired \textit{t}-test is used when comparing a pair of normally distributed sets of data subjected a single condition, here the effect of the training condition for each group.
To compare three or more groups of normally distributed data obtained from different sample populations, an Analysis of Variance (ANOVA) needs to be performed first to ascertain whether there are groups of data that are significantly different from other groups of data.
After ascertaining these differences, a post hoc \textit{t}-test is performed, whereby the pair-wise \textit{t}-test statistics are corrected to avoid minimise the occurrence of false positive results.
A repeated measure ANOVA is required when working with multiple sets of data collected for the same population, but subjected to different treatments.
A paired-sample \textit{t}-test was performed to compare the performance errors of baseline and evaluation phases for each group for both path following and waypoint navigation tasks.
To compare the differences between groups for the same phase (baseline and evaluation of each task), a one-way ANOVA followed by \textit{post-hoc} Holm-Sidak correction \textit{t}-test was employed.
The same method was used for the analysis of the path following training phase, where statistical comparisons were made between groups for specific amplitude or wavelength values.
For the same group, comparisons were performed for changing amplitude or wavelength values by carrying out a repeated measures ANOVA followed by a \textit{post-hoc} Holm-Sidak corrected \textit{t}-test. \cite{cardillo2008anovarep}.
Each of the significance tests were performed by assuming a significance level of $P$=0.05.
All statistical results are available in Data file S1, Supporting Information.


\medskip
\noindent \textbf{Supporting Information} \par 
\noindent Supporting Information is available from the Wiley Online Library or from the author.

\medskip
\noindent \textbf{Acknowledgements} \par 
\noindent The authors would like to thank Olexandr Gudozhnik for fabricating the electronic board used to power the electroadhesive clutches and his continued technical support over the course of the human subject experiments.  
We would also to thank Ronan Hinchet for his insightful comments to improve the performance of electroadhesive clutches.
We greatly appreciate the support of Enrico Ajanic whose endeavours have immensely benefited us in the writing of this article.
This work was partially funded by the European Union's Horizon  2020  research  and  innovation  program  under grant  agreement  ID:  871479  AERIAL-CORE as well as the Swiss National Science Foundation through the Swiss National Centre of Competence in Research (NCCR) Robotics.

\medskip.

\bibliographystyle{MSP}
\bibliography{intelbib}



\end{document}